# A Novel ECG Denoising Scheme Using the Ensemble Kalman Filter

Sadaf Sarafan, *Graduate Student Member*, *IEEE*, Hoang Vuong, Daniel Jilani, Samir Malhotra, Michael P.H. Lau, Manoj Vishwanath, *Graduate Student Member*, *IEEE*, Tadesse Ghirmai, *Senior Member, IEEE*, and Hung Cao, *Senior Member, IEEE*

*Abstract—* **Monitoring of electrocardiogram (ECG) provides vital information as well as any cardiovascular anomalies. Recent advances in the technology of wearable electronics have enabled compact devices to acquire personal physiological signals in the home setting; however, signals are usually contaminated with high level noise. Thus, an efficient ECG filtering scheme is a dire need. In this paper, a novel method using Ensemble Kalman Filter (EnKF) is developed for denoising ECG signals. We also intensively explore various filtering algorithms, including Savitzky-Golay (SG) filter, Ensemble Empirical mode decomposition (EEMD), Normalized Least-Mean-Square (NLMS), Recursive least squares (RLS) filter, Total variation denoising (TVD), Wavelet and extended Kalman filter (EKF) for comparison. Data from the MIT-BIH Noise Stress Test database were used. The proposed methodology shows the average signal to noise ratio (SNR) of 10.96, the Percentage Root Difference of 150.45, and the correlation coefficient of 0.959 from the modified MIT-BIH database with added motion artifacts.**

## I. Introduction

According to the Centers for Disease Control (CDC), cardiovascular diseases have become the first leading cause of death of all deaths occurring in the United States in 2017 [1]. According to the European Health Network European Cardiovascular Disease Statistics 2017 edition, each year cardiovascular disease causes 3.9 million deaths in Europe and over 1.8 million deaths in the European Union (EU) [2]. Assessment of electrocardiogram (ECG) will provide vital information about a heart condition, and the existence of abnormalities or distress. However, ECG recordings, especially by portable devices are commonly contaminated by outside interferences referred to as '*noise artifacts*'. These noise artifacts are a conglomerate of common noises such as motion noise, baseline wander, powerline interference, just to name a few. Despite the recent advances in signal processing, there is still no efficient method for denoising biopotentials acquired by wearables, such as ECG.

The current global COVID-19 pandemic has challenged the hospital-centered healthcare system and presented huge demands on a person-centered care system. The developing field of healthcare Internet of Things (IoT) offers more accessibility and mobility of medical services outside of a hospital setting. As the healthcare IoT field develops, mobile ECG devices will become more ubiquitous and will need to accommodate for everyday tasks that bed-ridden patients cannot perform. However, ECG recordings are commonly distorted contaminated by noise artifacts. Therefore, the extraction of high-resolution ECG signals from noisy measurements is required.

Several methods have been proposed to filter ECG from the signal contaminated with undesired interferences; however, each has its advantages and limitations. These include methods using smoothing filters such as Savitzky-Golay filtering (SG) [3], Extended Kalman Filter (EKF) [4], Wavelet Denoising (WD) [3], Empirical Mode Decomposition (EMD) [5], Ensemble Empirical Mode Decomposition (EEMD) [6], adaptive filtering like Recursive Least Squares (RLS) and Normalized Least-Mean-Square (NLMS) [3], Total Variation Denoising (TVD) [7], sparsity [8], among others. One study on ECG signals with noise levels from 5 dB signal to noise ratio (SNR) to 45 dB SNR showed that the WD denoises better than the others. However, SG and the Adaptive filter like RLS and NLMS perform better in some mid-range SNR [3]. The EKF is also a promising method; however, the filter model is highly reliant on the underlying dynamics assumed for the ECG signal and not practical for nonlinear models in a realistic environment [4]. Zebin *et al.* reported that after decomposing the ECG signal using EMD and applying soft wavelet thresholding to the high-frequency components, the reconstructed signal was denoised more effectively than if only one method were applied [5]. Kumar *et al.* applied TVD successfully for detecting R-peak signals with long-pause, drifts, complexes QRS, smaller R peaks, and even noisy signal portions. However, TVD is still comparatively less accurate than other methods for detecting false-positive and false-negative [7].

In this paper, we propose and develop a novel algorithm based on the Ensemble Kalman Filter (EnKF) to remove noise in ECG signals. Our results demonstrated that the proposed algorithm is reliable and effective as it could provide genuine ECG signals under strong noise condition.

## II. Methods and Implementation

### A. Ensemble Kalman Filter

The Kalman Filter (KF) was initiated by R. E. Kalman for linear models and the noises involved are additive [9]. For nonlinear cases, EKF has been widely used, which is based on local linearization of the nonlinear model using the

*Research supported by the NSF CAREER Award #1917105 (H.C.) and the NIH R44 #OD024874 (M.P.H.L. and H.C.). *Corresponding author: Hung Cao*. (e-mail: hungcao@uci.edu).

S. Sarafan, D.Jilani, and H. Cao are with the Electrical Engineering Department, the Samueli School of Engineering, UC Irvine, Irvine, CA 92697 USA (e-mail: ssarafan@uci.edu).

H.Vuong , M.P.H. Lau and H. Cao are with Sensoriis, Inc, Edmonds, WA, 98026 (e-mail: mlau@sensoriis.com).

Tadesse Ghirmai is with the School of STEM, University of Washington, Bothell, WA 98011 (e-mail: tadg@uw.edu).

S. Malhotra and H. Cao are with the Biomedical Engineering Department, UC Irvine, CA 92697.

M. Vishwanath and H. Cao are with the Computer Science Department, UC Irvine, CA 92697.

Jacobian [10]. The EKF-based algorithms have a drawback that, at every instant of time, they approximate the posterior probability density of the parameter of interest by a Gaussian distribution. When the true posterior density is not Gaussian, Sequential Monte Carlo (SMC) filtering methods show superior performance over EKF methods. G. Evensen introduced the EnKF [11] which is an approximate filtering method that represents the distribution of the state with an ensemble of draws from that distribution. Suppose the unknown time-varying state vector of a dynamic state-space model is denoted by $x_n \in R^{D_x}$ where $n = 1, 2, \ldots, N$ represents time instants and $D_x$ represents the dimension of $x_n$, and its evolution is given by:

$$x_n = f_n(x_{n-1}) + u_n \quad (1)$$

where $f(.)$ represents a state function which, in general, is nonlinear, and $u_n$ denotes the state noise vector with a known probability density function (pdf). Furthermore, the observation equation of the state-space model is given by:

$$y_n = h_n(x_n) + v_n \quad (2)$$

where $y_n \in R^{D_y}$ denotes the measurement vector obtained at time $n$, $D_y$ represents the dimension of the vector $y_n$, and $v_n$ denotes the measurement noise vector whose pdf is assumed known. Given the state-space model (1) and (2), our objective is to make a sequential estimate of the evolution of the state vector $x_{1:n} = \{x_1, \ldots, x_n\}$ in real-time as the measurement vector denoted by $y_{1:n} = \{y_1, \ldots, y_n\}$ becomes available. EnKF computes the Kalman gain by approximating $P_{xy,n}$ and $P_{yy,n}$ using their corresponding sample covariances, $\hat{P}_{xy,n}$ and $\hat{P}_{yy,n}$. To do so, $N$ number of ensembles, $\{x^{(i)}_{n|n-1}\}_{i=1}^{N}$, are first drawn from the prior probability density of the state vector, $p(x_{n|n-1})$, which has the same probability distribution function as the state noise with a mean of $f(x^{(i)}_{n-1})$. Once the ensembles are generated, the sample covariances of the errors are computed as follows:

$$\hat{P}_{xy,n} = \frac{1}{N} \sum_{i=1}^{N} (x^{(i)}_{n|n-1} - \underline{x}_n)(y^{(i)}_{n|n-1} - \underline{y}_n)^T \quad (3)$$

$$\hat{P}_{yy,n} = \frac{1}{N} \sum_{i=1}^{N} (y^{(i)}_{n|n-1} - \underline{y}_n)(y^{(i)}_{n|n-1} - \underline{y}_n)^T \quad (4)$$

where $y^{(i)}_{n|n-1} = h(x^{(i)}_{n|n-1})$, $\underline{x}_n = \frac{1}{N}\sum_{i=1}^{N} x^{(i)}_{n|n-1}$
$\underline{y}_n = \frac{1}{N}\sum_{i=1}^{N} y^{(i)}_{n|n-1}$. Then, the Kalman gain is
approximated by: $\hat{K}_n = \hat{P}_{xy,n} (\hat{P}_{yy,n})^{-1}$ (5)
the ensembles of the state vector, $\{x^{(i)}_n\}$, are computed as

$$x^{(i)}_n = x^{(i)}_{n|n-1} + \hat{K}_n(y_n + v^{(i)}_n - y^{(i)}_{n|n-1}) \quad (6)$$

where $v^{(i)}_n$ are samples obtained from Gaussian distribution with mean $\underline{y}_n$ and covariance $Q_w$. Once the ensembles of the state vector are computed, the estimate of the state vector is obtained by taking the averages of ensembles as follows:

$\hat{x}_n = \frac{1}{N}\sum_{i=1}^{N} x^{(i)}_n$ (7). All these are illustrated in **Fig. 1**.

### B. State-Space Model of a Synthetic ECG

A dynamic ECG model was proposed by McSharry *et. al.* to generate synthetic ECG trajectory in 3D Cartesian coordinate

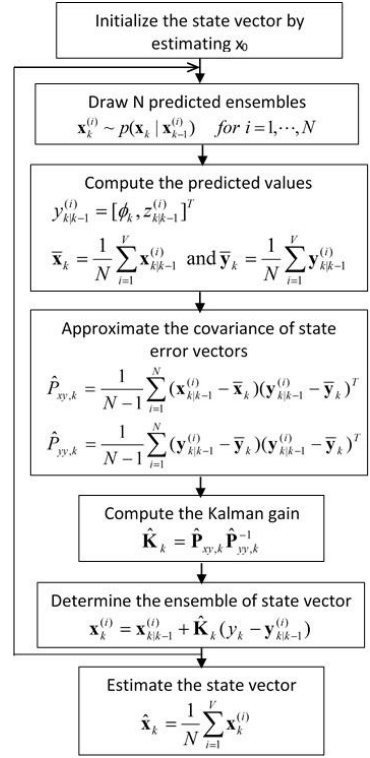

**Figure 1.** The EnKF algorithm flow chart.

[12]. Further, the ECG dynamic model is transformed into simpler polar form by Sameni *et al.* and provided a convenient discrete-time mathematical model [13]. The model represents an ECG signal by a sum of five Gaussian functions, each corresponding to the five waves of an ECG signal, namely P, Q, R, S, T waves. The state vector of the dynamic model is defined by $\mathbf{x}_k = [\theta_k, z_k]^T$, and the state equation is given by:

$$\begin{cases} \theta_k = (\theta_{k-1} + \omega.\Delta) \mod 2\pi \\ z_k = -\sum_{i \in [P,Q,R,S,T]} \frac{\alpha_i \Delta\theta_i \omega.\Delta}{b_i^2} \exp\left(-\frac{\Delta\theta_i^2}{2b_i^2}\right) + z_{k-1} + \eta_k \end{cases} \quad (8)$$

where $\Delta\theta_i = (\theta_k - \theta_i) \mod 2\pi$ is the phase increment, $\Delta$ is the sampling period, $\eta_k$ is the state noise, $\omega$ is the angular velocity of the trajectory as it moves around the limit cycle, and $\alpha_i$, $b_i$ and $\theta_i$ represent the amplitude, width, and center of the Gaussian functions, respectively. The measurement vector is defined by $\mathbf{y}_k = [\phi_k, s_k]^T$, where $\phi_k$ is the observed phase representing the linear time wrapping of the R-R time interval into $[0, 2\pi]$, and $s_k$ is the observed amplitude. The measurement equation is given by:

$$\begin{cases} \phi_k = \theta_k + u_k \\ s_k = z_k + v_k \end{cases}, \quad (9)$$

### C. Data for Testing

#### 1) Challenge databank

Generally, all algorithms under development would be evaluated using reliable and open online databanks. Our experiments use the MIT-BIH Noise Stress Test Database which includes twelve-half hours of ECG recordings and three-half hours of ECG with typical noise such as baseline wander, muscle artifact, and electrode motion artifact [14]. Clean ECG signals were used from the MIT-BIH Arrhythmia

Database (102, 108, 121, 122, 215, 220, 232, 118, and 119), and each dataset was calibrated at six noise levels based on different signal-to-noise ratios (SNR) from -6 to 24 dB at 360 samples per second.

### 2) Modified Signals with Motional Artifacts Added

When an ECG signal is recorded in daily life, it would be contaminated with many kinds of noise, such as motion artifacts. Unfortunately, the online dataset collected in the clinical setting is the ideal resting position with minimal motion artifacts. Therefore, here, we added the motion noise to the online clean dataset to have a better realistic scenario in our experiments. The ECG data were recorded using OpenBCI Cyton board (OpenBCI, Brooklyn, NY, USA) from healthy subjects during daily activities. Next, we normalized the recorded data to reinsure ECG and motion noise have realistic amplitudes. The EKF was employed to extract motion artifact noise. Then, ECG data should be normalized with the same threshold; before adding the motion noise to the new ECG. The process was previously described in [15].

### D. Comparison Criteria

To evaluate the performance of denoising algorithms, the improvement in SNR before and after denoising, the root mean square error (RMSE), the Percentage Root Difference (PRD), and the correlation coefficient between the denoised and the clean signal are calculated. Equations 10-13 gives the formulae for calculating these evaluations. where $x(n)$, and $y(n)$ are the original and the denoised signal respectively.

$$SNR = 10 \log \left( \frac{\sum_{i=1}^{n} x^2(n)}{\sum_{i=1}^{n} (x(n) - y(n))^2} \right) \quad (10)$$

$$RMSE = \sqrt{\frac{1}{n} \sum_{i=1}^{n} (x(n) - y(n))^2} \quad (11)$$

$$PRD = \sqrt{\frac{\sum_{i=1}^{n} (x(n) - y(n))^2}{\sum_{i=1}^{n} x^2(n)}} \times 100 \quad (12)$$

$$Corr = \frac{n(\sum_{i=1}^{n} x(n) y(n)) - [(\sum_{i=1}^{n} x(n))(\sum_{i=1}^{n} y(n))]}{\sqrt{\left[n \sum_{i=1}^{n} x(n)^2 - (\sum_{i=1}^{n} x(n))^2\right]\left[n \sum_{i=1}^{n} y(n)^2 - (\sum_{i=1}^{n} y(n))^2\right]}} \quad (13)$$

### III. EXPERIMENTS AND RESULTS

The three-dimensional trajectory which is generated from (8), consists of a unit circular (r=1) limit cycle which is going up and down when it approaches one of the P, Q, R, S, or T points. The projection of these trajectory points on the z-axis gives a synthetic ECG signal. We intensively explored various filtering algorithms, including EKF, SG, WD, EEMD, LMS, RLS, and TVD. **Fig. 2** illustrates the typical phase-wrapped results of the EnKF, EKF, SG, WD, EEMD, LMS, RLS, and the TVD for an input SNR of 12 dB. The signal under test is the record 118 from the MIT-BIH Arrhythmia Database with different SNR from 6 to 24 dB. The efficacy of the denoising techniques with the different input SNR levels is described in **Fig. 3**. The records from the MIT-BIH database are used. For each record, the SNR varies from 6 to 24 dB. This figure presents the SNR improvement, PRD, Correlation and RMSE for all the denoising techniques, at different levels of noise. **Table 1** presents the performance of all filtering algorithms performing on the modified MIT-BIH database with added motion artifacts.

### IV. DISCUSSION & CONCLUSION

In **Fig. 2**, results for filtering ECG performed by EnKF, EKF, SG, WD, EEMD, LMS, RLS, and TVD are depicted. As seen, the EnKF has followed the dynamics of the ECG signal and it has suppressed the noise more than the other methods. The proposed method, EnKF, is compared with those widely used filtering methods and it outperforms them in terms of visual quality. TVD and LMS annihilate the morphology of ECG which contains useful information. The SG, EEMD, and RLS could not eliminate the noise completely. The results of WD, and EKF are fairly consistent. **Fig. 3** illustrates the performance of different denoising methods at different input SNR levels. This figure compares RMSE, Correlation, PRD, and SNR of different denoising methods. **Fig. 3a** presents the comparison of the RMSE results obtained by using different denoising methods at the various level of SNR. As seen, for particular SNR levels, the proposed method yields the smallest RMSE thus demonstrates its capability to yield enhanced ECG signal with better quality. **Fig. 3b** depicts the correlation results for different input SNR levels. The proposed method has the maximum correlation with the clean signal. The proposed method provided significantly higher Correlation when compared with the other existing techniques. **Fig. 3c** indicates that the proposed denoising method provides lower PRD than other used methods. The mean of the SNR improvements versus different input SNRs is plotted in **Fig. 3d**. It is worth noting that EnKF has better performance in input SNRs greater than 10 dB. Visually comparing these results, it can be found the proposed methods have admirably tracked the original signal. This figure shows that the proposed method provided reasonably higher SNR in the higher SNR input level. To show that EnKF improvement is effective in daily life situations, the method has been validated using the modified data with motion noise added. **Table 1** shows the denoising performance comparison in terms of RMSE. Correlation, PRD, and SNR at the modified MIT-BIH database with added motion artifacts. It can be seen from the table that the EnKF has the highest SNR improvement. The same conclusion can be drawn from the PRD and Correlation.

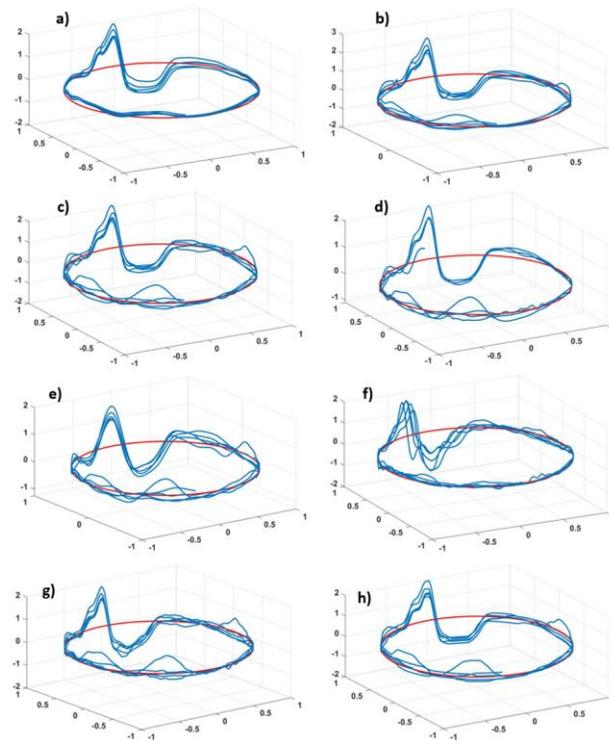

**Figure 2.** Phase-wrapped ECG filtering results for an input signal of 12 dB. (**a**) EnKF. (**b**) EKF. (**c**) SG. (**d**) Wavelet denoising. (**e**) EEMD filter. (**f**) LMS adaptive filter. (**g**) RLS adaptive filter. (**h**) Total variation denoising.

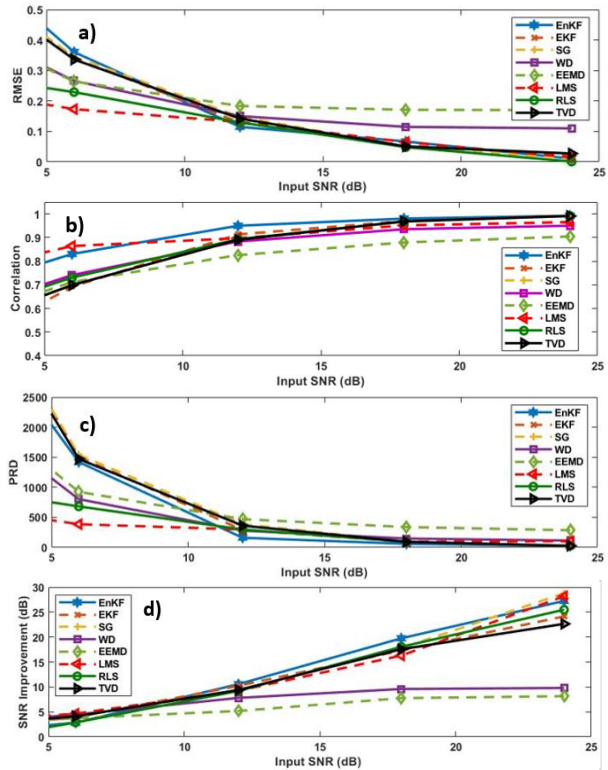

**Figure 3.** (**a**) Comparison of the mean RMSE obtained by using different denoising methods at different input SNR levels. (**b**) Comparison of the mean Correlation versus different input SNR levels. (**c**) Comparison of the PRD for different denoising methods at different input SNR levels. (**d**) Comparison of the mean SNR improvements for different denoising methods versus different input SNR levels.

TABLE I. Performance of Denoising Algorithms for the Motion Noise Cancellation

|  | *SNR* | *Correlation* | *PRD* | *RMSE* |
|---|---|---|---|---|
| EnKF | 10.965 | 0.959 | 150.45 | 0.105 |
| EKF | 10.226 | 0.913 | 288.18 | 0.128 |
| SG | 9.280 | 0.888 | 387.12 | 0.148 |
| WD | 7.813 | 0.883 | 277.82 | 0.150 |
| EEMD | 5.187 | 0.825 | 469.50 | 0.183 |
| LMS | 9.254 | 0.897 | 297.12 | 0.131 |
| RLS | 9.206 | 0.892 | 298.15 | 0.129 |
| TVD | 9.397 | 0.890 | 361.32 | 0.142 |

The results of this paper approve the applicability of the EnKF, for the filtering of noisy ECG signals. In the future, we would like to optimize our denoising method and test with raw ECG data. We are developing a novel wearable ECG monitoring system in real-time and pseudo-real time combined with our denoising method EnKF [16]. The device will provide analytics details on our mobile Android and iOS apps using cloud computing.

REFERENCES

[1] M. P. Heron, "Deaths: leading causes for 2017," 2019.
[2] E. Wilkins *et al.*, "European cardiovascular disease statistics 2017," 2017.
[3] M. AlMahamdy *et al.*, "Performance study of different denoising methods for ECG signals," *Procedia Computer Science,* vol. 37, pp. 325-332, 2014.
[4] R. Sameni *et al.*, "Filtering electrocardiogram signals using the extended Kalman filter," in *2005 IEEE Engineering in Medicine and Biology 27th Annual Conference*, 2006: IEEE, pp. 5639-5642.
[5] L. Zebin *et al.*, "Research on ECG Denoising method Based on Empirical Mode Decomposition and Wavelet Transform," in *2020 IEEE 5th International Conference on Signal and Image Processing (ICSIP)*, 2020: IEEE, pp. 675-679.
[6] K.-M. Chang, "Arrhythmia ECG noise reduction by ensemble empirical mode decomposition," *Sensors,* vol. 10, no. 6, pp. 6063-6080, 2010.
[7] S. S. Kumar, *et al.*, "Total variation denoising based approach for R-peak detection in ECG signals," *Procedia Computer Science,* vol. 93, pp. 697-705, 2016.
[8] N. Mourad, "ECG denoising algorithm based on group sparsity and singular spectrum analysis," *Biomedical Signal Processing and Control,* vol. 50, pp. 62-71, 2019.
[9] R. E. Kalman, "A new approach to linear filtering and prediction problems," *Journal of basic Engineering,* vol. 82, no. 1, pp. 35-45, 1960.
[10] A. H. Jazwinski, *Stochastic processes and filtering theory*. Courier Corporation, 2007.
[11] G. Evensen, "Sequential data assimilation with a nonlinear quasi-geostrophic model using Monte Carlo methods to forecast error statistics," *Journal of Geophysical Research: Oceans,* vol. 99, no. C5, pp. 10143-10162, 1994.
[12] P. E. McSharry, G. D. Clifford, L. Tarassenko, and L. A. Smith, "A dynamical model for generating synthetic electrocardiogram signals," *IEEE transactions on biomedical engineering,* vol. 50, no. 3, pp. 289-294, 2003.
[13] R. Sameni *et al.*, "A nonlinear Bayesian filtering framework for ECG denoising," *IEEE Transactions on Biomedical Engineering,* vol. 54, no. 12, pp. 2172-2185, 2007.
[14] G. B. Moody *et al.*, "A noise stress test for arrhythmia detectors," *Computers in cardiology,* vol. 11, no. 3, pp. 381-384, 1984.
[15] S. Sarafan, *et al.*, "Investigation of Methods to Extract Fetal Electrocardiogram from the Mother's Abdominal Signal in Practical Scenarios," *Technologies,* vol. 8, no. 2, p. 33, 2020.
[16] S. Sarafan, *et al.,* "Development of a Home-based Fetal Electrocardiogram (ECG) Monitoring System." *2021 43rd Annual International Conference of the IEEE Engineering in Medicine & Biology Society (EMBC)*. IEEE, 2021.